\documentclass[runningheads,citeauthoryear]{apinv}
\usepackage{epsfig,cite,graphics}
\usepackage[english]{babel}

\begin{document}

\title{3C\,273~-- half a century later}
\titlerunning{3C\,273~-- half a century later}
\author{L. Slavcheva-Mihova, B. Mihov, I. Iliev}
\authorrunning{L. Slavcheva-Mihova et al.}
\tocauthor{L. Slavcheva-Mihova, B. Mihov, I. Iliev}
\institute{Institute of Astronomy and NAO, Bulgarian Academy of Sciences, BG-1784, Sofia
\newline
\email{lslav@astro.bas.bg}}
\papertype{Research report. Accepted on xx.xx.xxxx}	
\maketitle

\begin{abstract}
We have presented an optical monitoring of 3C\,273, the first quasar discovered fifty years ago. It does not show variability both on intra-night and long-term time scales. To facilitate the further monitoring of 3C\,273, we compiled the available calibrations of the comparison stars in its field into a mean sequence.
\end{abstract}

\section{Introduction}
\label{intro}
Bright enough in the optical, 3C\,273 was known from old survey plates at least as early as 1887. It was first catalogued in the Third Cambridge Catalogue of Radio Sources (3C, Edge et al. 1959). Fifty years ago, upon his attempts to identify the emission lines in the optical spectrum of this radio-loud 
star-like object of thirteenth magnitude, Maarten Schmidt discovered the most powerful long-lived phenomena in the Universe, active galactic nuclei (AGNs, Schmidt 1963).

The AGN model involves accretion around a supermassive black hole. Blazars are jet-on AGNs. They are variable at all frequencies, especially at high energies with minimum variability timescales down to tens of minutes. Variability may be coordinated in different energy bands. 
Blazars could be divided into BL\,Lacs with absent or weak emission lines with equivalent width $\rm EW\!<\!5\,\AA$ and flat-spectrum radio quasars (FSRQs) showing strong emission lines with $\rm EW\!>\!5\,\AA$. 

3C\,273 is a FSRQ. Apart from being the first quasar discovered, it is optically the brightest quasar with $V\!=\!12.85$\,mag and among the nearest ones with a redshift of $z\!=\!0.158$ (V\'eron-Cetty \& V\'eron 2010).

Right after revealing the true nature of 3C\,273, its brightness was followed on archival photographic plates back to the end of the 19$^{\rm th}$ century (Smith 1965; Angione \& Smith 1985)~-- the quasar showed variations up to 1.5\,$B$\,mag. Flux changes of a similar amplitude were detected in the recent CCD light curves (e.g., Fan et al. 2009b), too. Analyzing the $B$ band light curve during 114 years, Dai et al. (2006) found
a variability period of 13.51\,yr. At the present time, the most complete long-term light curves of 3C\,273 could be found in the database available at the ISDC\footnote{http://isdc.unige.ch/3c273/} Data Centre for Astrophysics, Switzerland (T\"urler et al. 1999; Soldi et al. 2008).

Regarding the intra-night time scales, 3C\,273 seems not very active in the optical. The first hint for microvariability was reported by Moles et al. (1986)~-- the $I$ band magnitude changed by 0.19\,mag in less than 2\,h. Ghosh et al. (2000), Dai et al. (2001), and Romero et al. (2002) did not find intra-night 
variability in a total of five attempts. Dai et al. (2009) did not detect microvariability in their 2003--2005 campaign except on one occasion, which they found not convincing. Fan et al. (2009a) detected $R$ band variations of 0.18\,mag over 2.1\,h. According to Rani et al. (2011), the source showed microvariability in two out of eight nights. 

Soon after the key paper of Schmidt (1963), a photographic comparison star sequence in the field of 3C\,273 was introduced (Sharov \& Efremov 1963; Smith \& Hoffleit 1963). Calibrations based on photoelectric (Smith 1965; Burkhead \& Parvey 1968; Penston et al. 1971; Smith et al. 1985; Hamuy \& Maza 1989) and CCD (Villata et al. 1997; Fiorucci et al. 1998; Gonz\'alez-P\'erez et al. 2001; Doroshenko et al. 2005) data followed.
The transformation to the standard Johnson-Cousins system was done using Landolt stars (Smith et al. 1985; Villata et al. 1997; Fiorucci et al. 1998; Gonz\'alez-P\'erez et al. 2001); stars of spectral type near A0 (Burkhead \& Parvey 1968); primary standards from Johnson \& Morgan (1953) list and other secondary standards (Penston et al. 1971); stars listed in Graham (1982; Hamuy \& Maza  1989); the E star from Gonz\'alez-P\'erez et al. (2001; Doroshenko et al. 2005). 

The aims of the present work are to study 3C\,273 intra-night variability and its relation to the long-term variations. We would also like to combine various calibrations of the comparison stars into a mean sequence to facilitate the further monitoring of the quasar.

The paper is organized as follows. In Sect.\,\ref{ss} we construct a mean comparison sequence in the field of 3C\,273. The intra-night monitoring is presented in Sect.\,\ref{inom}. Discussion of the results follows in Sect.\,\ref{disc}.

\section{Mean comparison sequence}
\label{ss}

\begin{figure}[!htp]
\begin{center}
\centering{\epsfig{file=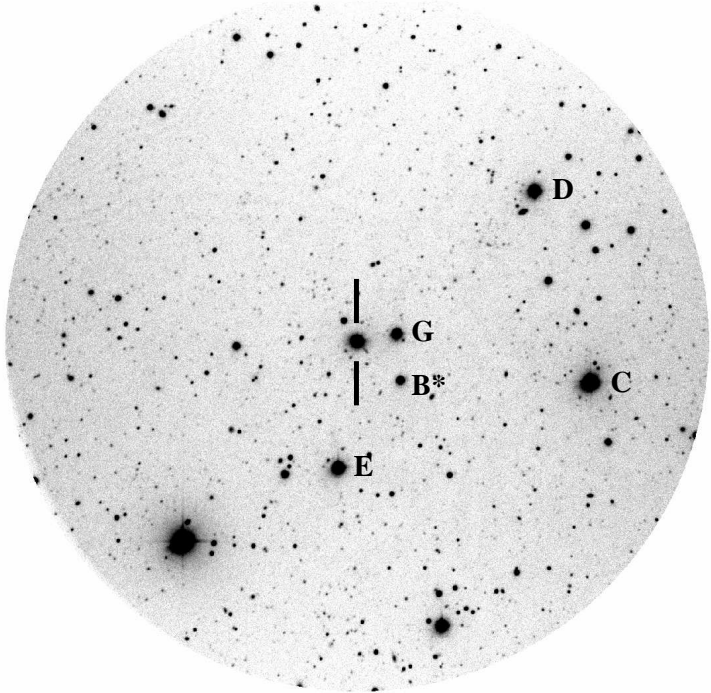, width=0.75\textwidth}}
\caption[]{Comparison stars in the field of 3C\,273, as denoted in Penston et al. (1971). This $R$ band image is an average of three 300\,s frames taken during the March monitoring (see Sect.\,\ref{inom})~-- the quasar jet is clearly visible. North is at the top, east to the left.}
\label{f_stds}
\end{center}
\end{figure}

\begin{figure}[!htp]
\begin{minipage}[t]{0.48\linewidth}
\centering
\includegraphics[width=\textwidth]{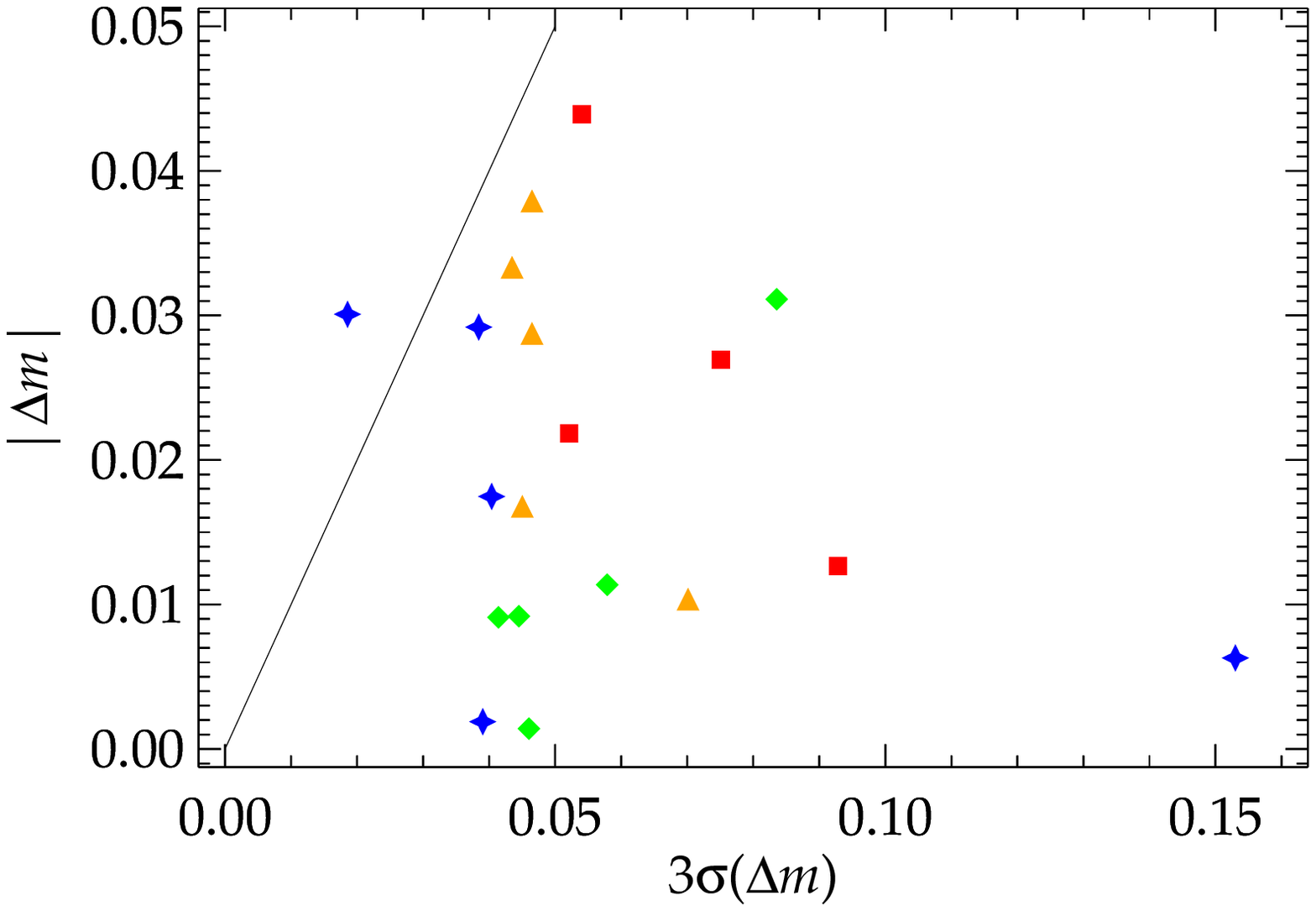}
\caption[]{Comparison between the absolute difference of the magnitude (of the photoelectric and CCD data) and three times the uncertainty of the difference. Blue stars, green diamonds, orange triangles, and red squares mark the $BVRI$ bands, respectively.}
\label{el_CCD}
\end{minipage}
\hfill
\begin{minipage}[t]{0.48\linewidth}
\centering
\includegraphics[width=\textwidth]{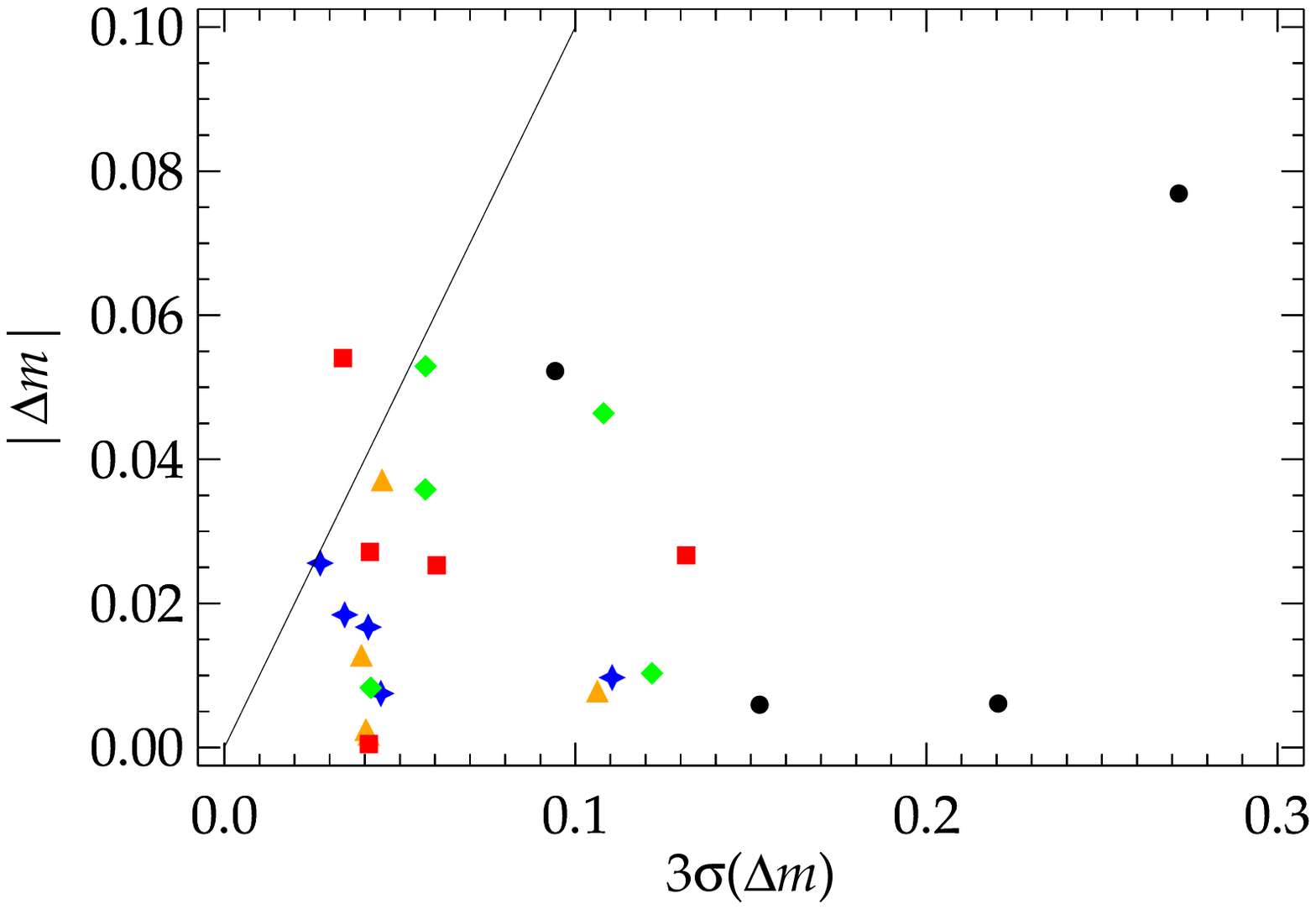}
\caption[]{Comparison between the absolute difference of the magnitude (of the calibrations using Landolt stars and stars from other lists, see text) and three times the uncertainty of the difference. Black dots, blue stars, green diamonds, orange triangles, and red squares mark the $UBVRI$ bands, respectively.}
\label{Land_oth}
\end{minipage}
\end{figure}

We checked for systematic offsets between the calibrations of the comparison sequence based on photoelectric and CCD data; the stars in common are shown in Fig.\,\ref{f_stds}. We weight-averaged the two sets of data (except those of Smith 1965 due to a lack of magnitude uncertainties) on an individual basis for each of the stars in common and each band. The outliers were removed after Odewahn et al. (1992).
In Fig.\,\ref{el_CCD} the differences between the photoelectric and CCD magnitudes against the uncertainties of the differences are plotted.
A similar check was performed for offsets between calibrations based on Landolt stars and other standards (Fig.\,\ref{Land_oth}). The lack of systematic differences can be seen~-- in both cases all values but one lie within the $3\,\sigma$ range. 

Thus, the individual calibration sets show good agreement among themselves without systematic dependence on the detector type or standard stars used for calibration. This made us weight-average the photoelectric (except the data of Smith 1965) and CCD data altogether with outliers not taken into account for the stars in common to produce a mean comparison sequence around 3C\,273 (Fig.\,\ref{f_stds}).
The number of the data points averaged, $N$, the weight-averaged magnitudes, $\langle m \rangle$, their uncertainties, $\sigma(\langle m \rangle)$, and the resulting $\chi^2_{\rm df}$ are listed in Table\,\ref{t_stds}.

\begin{table}[t]
\begin{center}
\caption{Weight-averaged Johnson-Cousins magnitudes of the comparison stars in the field of 3C\,273.}
\begin{tabular}{l@{\hspace{1cm}}l@{\hspace{1cm}}l@{\hspace{1cm}}r@{\hspace{1cm}}r@{\hspace{1cm}}r}
\hline\noalign{\smallskip}
Band & Star & $N$ & $\langle m \rangle$ & $\sigma(\langle m \rangle)$ & $\chi^2_{\rm df}$ \\
\noalign{\smallskip}\hline\noalign{\smallskip}
   $U$ & B* &  2  &  17.009  &  0.018  &  1.94 \\
       &  C &  3  &  13.422  &  0.009  &  0.24 \\
       &  D &  4  &  13.144  &  0.009  &  0.09 \\
       &  E &  3  &  13.467  &  0.011  &  0.86 \\
       &  G &  3  &  14.231  &  0.011  &  0.43 \\ 
\hline\noalign{\smallskip}
   $B$ & B* &  3 &  16.025 &  0.007 &  0.86 \\
       &  C &  4 &  12.874 &  0.010 &  0.16 \\
       &  D &  7 &  13.165 &  0.006 &  1.52 \\
       &  E &  5 &  13.364 &  0.004 &  1.03 \\
       &  G &  6 &  14.125 &  0.002 &  0.64 \\
\hline\noalign{\smallskip}
   $V$ & B* &  4 &  14.924  & 0.007 &  0.62 \\
       &  C &  4 &  11.880  & 0.007 &  0.02 \\
       &  D &  4 &  12.648  & 0.007 &  0.53 \\
       &  E &  5 &  12.712  & 0.007 &  0.17 \\
       &  G &  6 &  13.538  & 0.006 &  0.80 \\
\hline\noalign{\smallskip}
   $R$ & B* &  3 &  14.212 &  0.007 &  0.09 \\
       &  C &  4 &  11.281 &  0.010 &  2.01 \\
       &  D &  4 &  12.308 &  0.002 &  0.50 \\
       &  E &  5 &  12.287 &  0.006 &  1.70 \\
       &  G &  6 &  13.173 &  0.005 &  1.06 \\
\hline\noalign{\smallskip}
   $I$ & B* &  2 &  13.599 &  0.010 &  0.0004 \\
       &  C &  2 &  10.748 &  0.016 &  0.05 \\
       &  D &  3 &  11.988 &  0.008 &  1.02 \\
       &  E &  3 &  11.901 &  0.009 &  0.80 \\
       &  G &  4 &  12.844 &  0.006 &  0.13 \\
\hline\noalign{\smallskip}
\end{tabular}
\label{t_stds}
\end{center}
\end{table}

\section{Intra-night optical monitoring}
\label{inom}

The intra-night monitoring\footnote{The light curves could be found at www.astro.bas.bg/$\sim$bmihov.} of 3C\,273 was performed on March 10 and April 14, 2013. During the first night we used the 2\,m telescope of the Rozhen National Astronomical Observatory (NAO), Bulgaria,
equipped with a two-channel focal reducer and a $1340\,\times\,1300$ Princeton Instruments VersArray 1300B CCD camera (a pixel size of $20\,\mu\rm m$ that gives a scale factor of 0.737 arcsec/px). The exposure times were 30/35/40\,s through the Cousins $R$ filter; three 300\,s images were also obtained.
During the second night we used the 50/70\,cm Schmidt telescope of NAO with a $4096\!\times\!4096$ FLI PL16803 CCD camera (a pixel size of $9\,\mu\rm m$ that gives a scale factor of 1.079\,arcsec/px). The exposure times were 180\,s through the $B$ filter and 90\,s through the Cousins $R$ filter.
In addition, on Febr. 11, 2013 two data sets (in the $VR$ bands with exposure times ranging from 60\,s to 300\,s) were obtained with 2.9\,h in between them; the observational setup was the same as for March 10. 

The primary reduction was done in the trivial way: bias (for the 2\,m telescope data) and dark current (for the Schmidt telescope data) subtraction followed by flat fielding and cosmic ray hits removal. The photometry was performed using DAOPHOT (Stetson 1987). We used a $2\!\times\!FWHM$ aperture
radius (but not smaller than 7\arcsec\ for the Schmidt telescope data to avoid unreasonably small aperture sizes). The stars listed in Table\,\ref{t_stds} were used as comparison ones; star C was not used for the March photometry as it got saturated. Star E was used as a control one.

To test 3C\,273 for intra-night variability, we used the scaled version of the $F$ statistics (see Howell et al. 1988 for details)~-- the scaling prevents us from a false variability detection (non-detection) due to the different brightness between the target object and the control star. Our choice of star E as a control one results in a scale factor close to unity (see Table\,\ref{t_inom}, col.\,9). The $\chi^2$ statistics could be used as an alternative variability test (de Diego 2010).

\begin{figure}[!htp]
\begin{minipage}[t]{0.48\linewidth}
\centering
\includegraphics[width=\textwidth]{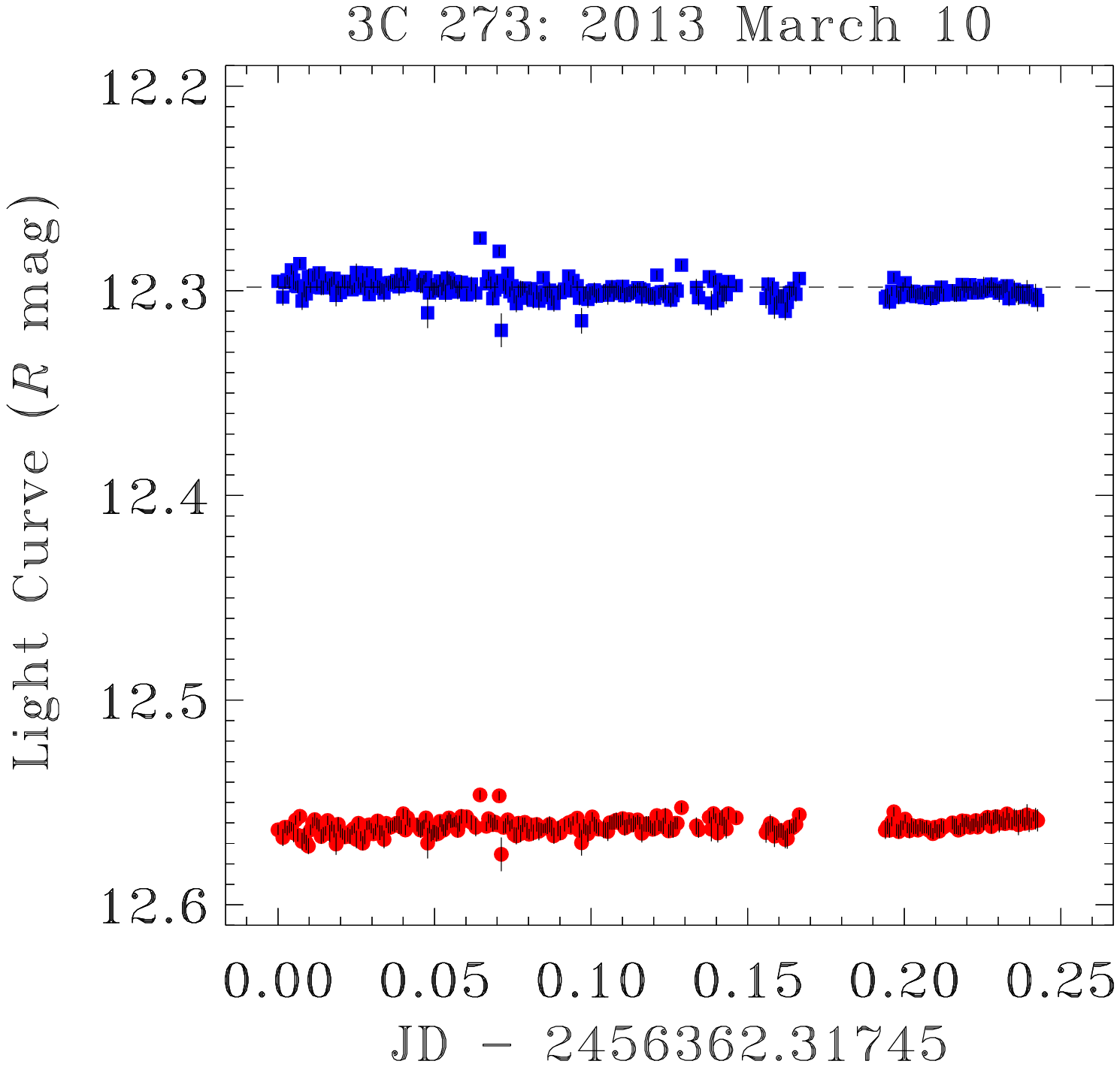}
\caption[]{Intra-night light curve of 3C\,273 (red circles) and of star E (blue squares) obtained with the 2\,m telescope in the $R$ band; the dashed line denotes the weight-averaged magnitude of the control star.}
\label{lc_r_2m}
\end{minipage}
\hfill
\begin{minipage}[t]{0.48\linewidth}
\centering
\includegraphics[width=\textwidth]{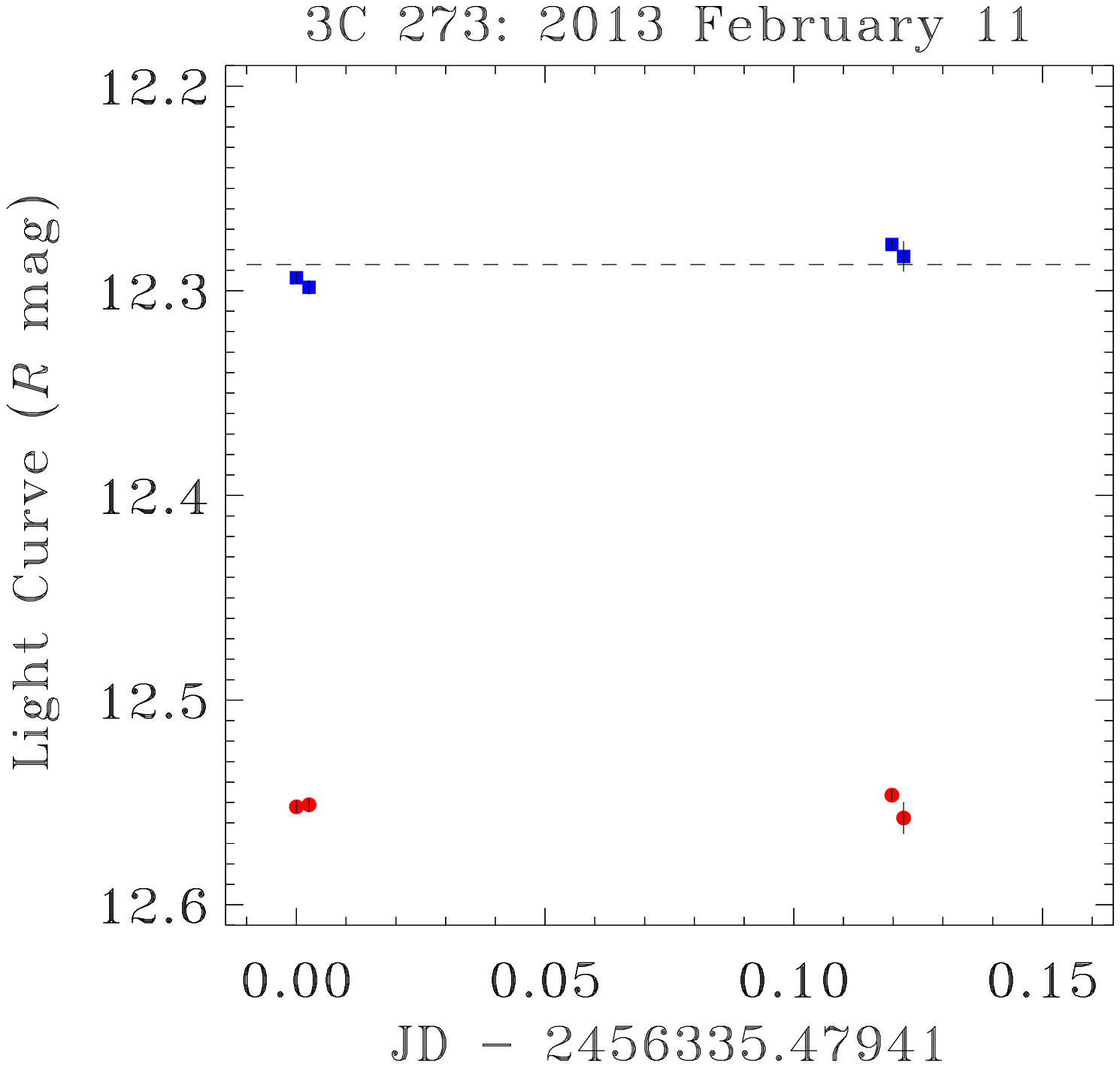}
\caption[]{The same as in Fig.\,\ref{lc_r_2m}.}
\label{lc_rf}
\end{minipage}
\end{figure}

\begin{figure}[!htp]
\begin{minipage}[t]{0.48\linewidth}
\centering
\includegraphics[width=\textwidth]{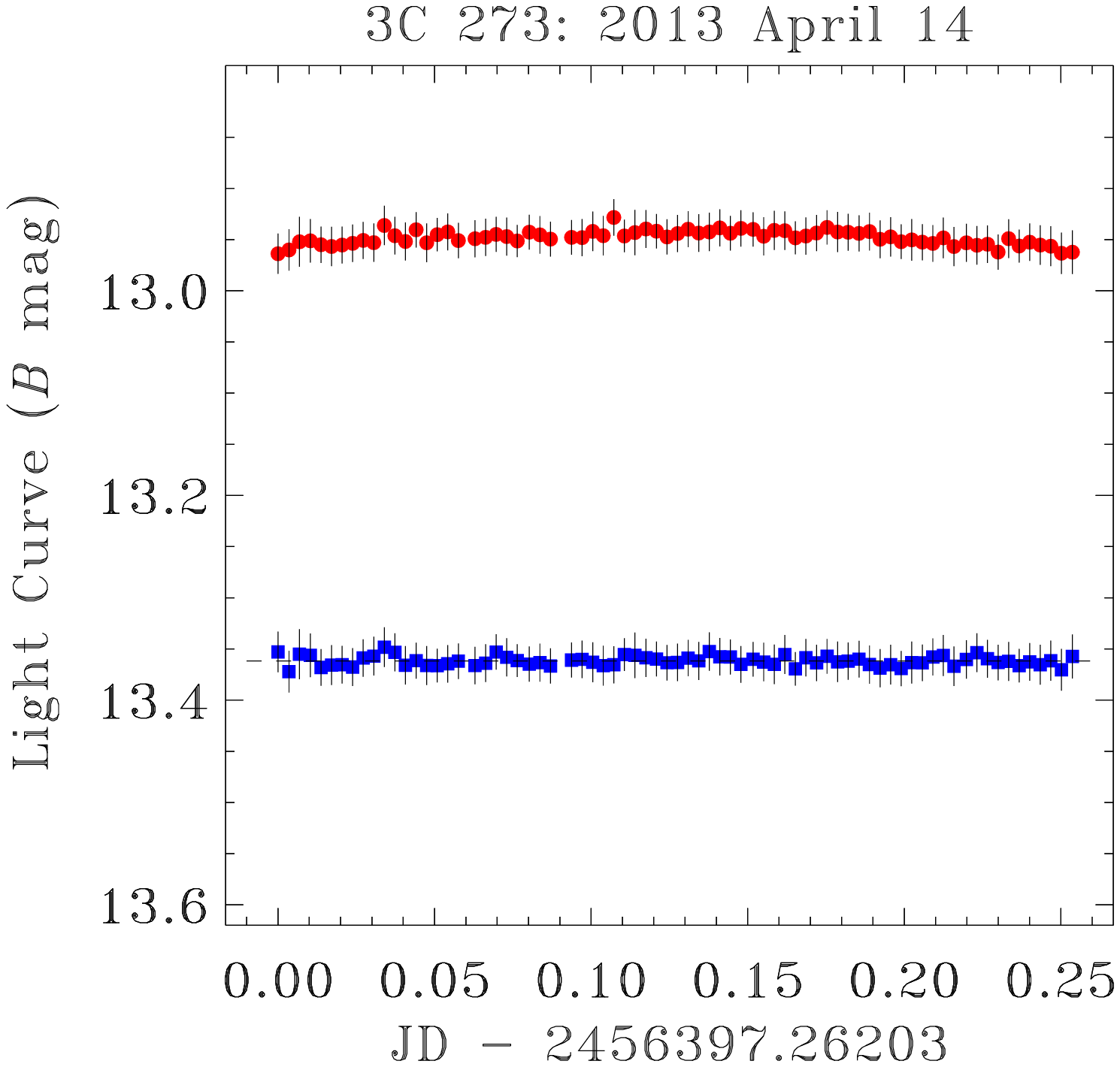}
\caption[]{The same as in Fig.\,\ref{lc_r_2m} but for the Schmidt telescope in the $B$ band.}
\label{lc_b_sch}
\end{minipage}
\hfill
\begin{minipage}[t]{0.48\linewidth}
\centering
\includegraphics[width=\textwidth]{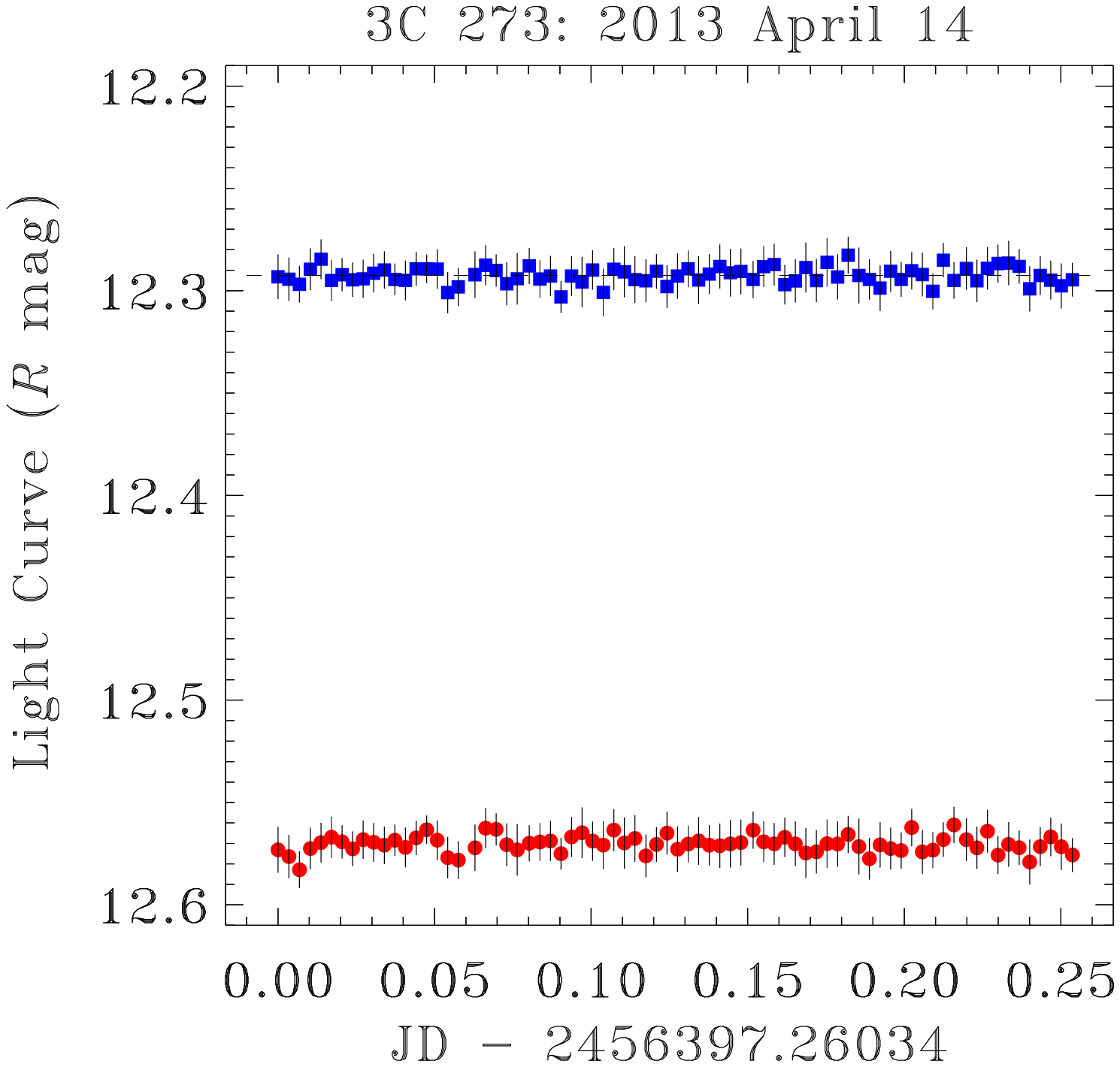}
\caption[]{The same as in Fig.\,\ref{lc_r_2m} but for the Schmidt telescope in the $R$ band.}
\label{lc_r_sch}
\end{minipage}
\end{figure}

The results from the intra-night monitoring of 3C\,273 in 2013 are shown in Figs.\,\ref{lc_r_2m},\,\ref{lc_rf},\,\ref{lc_b_sch},\,\ref{lc_r_sch} and summarized in Table\,\ref{t_inom}. The columns read the following: $\Delta t$ is the monitoring duration, $N_{\rm pts}$ the number of  data points, $\langle m \rangle$ the weight-averaged magnitude of the quasar (col.\,5) and of the control star (col.\,6, the value of the resulting $\chi^2_{\rm df}$ is
also given), $\sigma$ denotes the weighted standard deviation of the corresponding light curves about the weight-averaged magnitudes, med($\Gamma^2$) is the median value of the $\Gamma^2$ factor that accounts for the different brightness between the quasar and the control star (see Howell et al. 1988), 
$F$ denotes the calculated (col.\,10) and critical (for $\alpha\!=\!0.05$, col.\,11) values of the $F$ statistics, and the last column indicates the presence of variability. We got $V\!=\!12.83\!\pm\!0.03$ for the quasar and $V\!=\!12.73\!\pm\!0.03$ for star E. The magnitudes of star E obtained in the course of the monitoring are in good agreement with the values presented in Table\,\ref{t_stds}. We obtain the following colour indices for 3C\,273: $B\!-\!R\!=\!0.378\!\pm\!0.002$ and $V\!-\!R\!=\!0.28\!\pm\!0.03$.

\begin{table}[t]
\begin{center}
\caption{Results from the intra-night monitoring of 3C\,273 during 2013. Uncertainties of the weight-averaged
magnitudes are given in parentheses and are in units of 0.0001\,mag.}
\begin{tabular}{l@{\hspace{.5cm}}l@{\hspace{.5cm}}l@{\hspace{.5cm}}l@{\hspace{.5cm}}r@{\hspace{.5cm}}r}
\hline\noalign{\smallskip}
Month & Filter & $\Delta t$ [h]& $N_{\rm pts}$ & $\langle m(\rm 3C\,273) \rangle$, $\chi^2_{\rm df}$ & $\langle m(\rm star\,E) \rangle$, $\chi^2_{\rm df}$ \\
\noalign{\smallskip}\hline\noalign{\smallskip}
(1) & (2) & (3) & (4) & (5) & (6) \\
\hline\noalign{\smallskip}
February & $R$ & 2.9 & 4 &  12.5498 (18), 1.2 & 12.2872 (52), 10.9 \\
\noalign{\smallskip}
March & $R$ & 5.8 & 276 & 12.5610 (2), 1.2 & 12.2982 (3), 2.1 \\
\noalign{\smallskip}
April & $B$ & 6.1 & 74  & 12.9480 (21), 0.1 & 13.3618 (21), 0.1 \\
\noalign{\smallskip}
April & $R$ & 6.1 & 75  & 12.5701 (11), 0.2 & 12.2925 (11), 0.2 \\
\hline\noalign{\smallskip}
$\sigma[m(\rm 3C\,273)]$ & $\sigma[m(\rm star\,E)]$ & med($\Gamma^2$) & $F_{\rm calc}$ & $F_{\rm crit}(\alpha=0.05)$ & Variability \\
\noalign{\smallskip}\hline\noalign{\smallskip}
(7) & (8) & (9) & (10) & (11) & (12) \\
\hline\noalign{\smallskip}
0.0036 & 0.0103 & 1.164 & 0.198 & 9.277 & no \\
\noalign{\smallskip}
0.0034 & 0.0044 & 1.084 & 0.577 & 1.220 & no \\
\noalign{\smallskip}
0.0067 & 0.0047 & 0.870 & 2.287 & 1.473 & yes \\
\noalign{\smallskip}
0.0043 & 0.0042 & 1.112 & 0.929 & 1.469 & no \\
\hline\noalign{\smallskip}
\end{tabular}
\label{t_inom}
\end{center}
\end{table}

\section{Discussion}
\label{disc}

According to the $F$ statistics, 3C\,273 does not show intra-night variability except in the $B$ band (see Table\,\ref{t_inom}). However, the small value of $\chi^2_{\rm df}$ in this case prompts
variations within the formal uncertainties of the measurements. Moreover, there are no 
indications for variability in the $R$ band for the same night. Therefore, we consider the variations of the 3C\,273 light curve in the $B$ band not caused by the quasar variability.

The weight-averaged quasar magnitudes in the period Febr.--April show a weak fading tendency (a gradient of $0.11\!\pm\!0.01$\,mag/yr). The brightness, however, changes by about 0.02\,mag hinting an inactive state. For that period we could find only an unpublished $R$ band light curve obtained in the course 
of the Tuorla quasar monitoring\footnote{http://users.utu.fi/kani/1m/3C\_273.html} (Katajainen et al. 2000, present are only graphical data) and the rough $V$ band photometry obtained by the Catalina Surveys\footnote{http://nesssi.cacr.caltech.edu/catalina/Blazars/501021010674100113p.html} (Drake et 
al. 2009). In both data sets the quasar does not show violent activity, $\chi^2_{\rm df}\!=\!0.43$ for the Catalina data. The fading tendency could be traced on the Tuorla data, too. In conclusion, based on the available data, we found that in the first months of 2013 the quasar is in a quiescent phase on both intra-night and long-term time scales.

Howard et al. (2004), analyzing the variability data for four blazars, found evidence that the intra-night variability occurs more frequently when the 
flux is changing on longer time scales. Similar claim was made by Dultzin-Hacyan et al. (1997) for OJ\,287. Our results are in a general agreement with this scenario. 

\section{Summary}
\label{sum}

In this paper we have presented an intra-night optical monitoring of 3C\,273. The quasar does not show variability. We did not find indications for variability on a long-term time scale either, inspecting both our data and archives for the period Febr.--April, 2013.
The published calibrations of the comparison stars in common in the photoelectric and CCD data in the field of 3C\,273 were combined. Thus, we compiled a mean comparison sequence of five stars measured in five bands.

\section*{Acknowledgements}
The CSS survey is funded by the National Aeronautics and Space
Administration under Grant No. NNG05GF22G issued through the Science
Mission Directorate Near-Earth Objects Observations Program.  The CRTS
survey is supported by the U.S.~National Science Foundation under
grants AST-0909182.

\end{document}